\documentclass[epj]{svjour}
\usepackage[utf8]{inputenc}
\usepackage{graphicx}
\usepackage{flafter}

\usepackage{amsmath,amssymb,amstext,amsfonts}
\usepackage{braket}
\usepackage{bm}
\usepackage{float}
\usepackage{color}
\usepackage{subfigure}
\usepackage{blindtext}
\usepackage{marginnote}

\let \Re \relax
\DeclareMathOperator{\Re}{Re}
\let \Im \relax
\DeclareMathOperator{\Im}{Im}

\begin{document}


\title{Resonant scattering of Dice quasiparticles on oscillating quantum dots}

\author{Alexander Filusch\thanks{alexander.filusch@uni-greifswald.de}, Christian Wurl\thanks{chr.wurl89@gmail.com} \and Holger Fehske\thanks{fehske@physik.uni-greifswald.de}}

\institute{Institut f{\"u}r Physik,
Universit{\"a}t Greifswald, 17487 Greifswald, Germany }


\date{\today}

\abstract{
We consider a Dice model with Dirac cones intersected by a topologically flat band  at the charge neutrality point and analyze the inelastic scattering of massless pseudospin-1 particles on a circular, gate-defined,  oscillating barrier. 
Focusing on the resonant scattering regime at small energy of the incident wave, we calculate the reflection and transmission coefficients and derive explicit expressions for the time-dependent particle probability, current density and scattering efficiency within (Floquet) Dirac-Weyl theory,  both in the near-field and the far-field.  
  We discuss the importance of sideband scattering and Fano resonances in the quantum limit. When resonance conditions are fulfilled, the particle is temporarily trapped in vortices located close to edge of the quantum dot before it gets resubmitted with strong angular dependence. Interestingly even periodically alternating forward and backward radiation may occur.  We also demonstrate the revival of resonant scattering related to specific fusiform boundary trapping profiles.
}
\authorrunning{A. Filusch et al.}
\maketitle

\section{Introduction}
Solid states systems dominated by Dirac-cone physics constitute a unique form of quantum matter being attractive for both fundamental science and technology.  The discovery of such massless chiral Dirac fermions in graphene~\cite{CGPNG09}, on the surface of topological insulators ~\cite{HK10}, and in Weyl semimetals~\cite{Xu15} have boosted progress towards the theoretical modelling of these materials in the last decade. Their striking and sometimes counterintuitive electronic, spectroscopic and transport properties primarily arise from the linear (gapless) energy spectrum near Dirac nodal points, pseudospin conservation  and the related topology of the wave function. In this context, one of the most spectacular findings was the direct experimental observation of Klein tunneling~\cite{Kl28} in graphene~\cite{KNG06,SHG09,YK09}, i.e., of a perfect transmission of quasirelativistic pseudospin-1/2 particles through potential barriers, which seems to prevent any electrostatic confinement of Dirac-Weyl electrons.

Around the same time, the transmission properties of  massless pseudospin-1 particles have also attracted much interest~\cite{BUGH09}. Such Dirac-Weyl quasiparticles can be viewed as low-energy excitations of the Dice lattice model having an additional atom at the center of the hexagons of the honeycomb graphene lattice. In the Dice system, a dispersionless band---related to strictly localized states in view of the local topology~\cite{Su86}---crosses the conical connecting points with surprising consequences.  For example, electrostatic barriers become even more transparent compared to the pseudospin-1/2 system and there is a regime of angle-independent super-Klein-tunneling~\cite{UBWH11}. In general, flat band systems establish exceptional phases of matter,  and consequently have been engineered not only for electrons, but also for cold atoms and photons~\cite{LAF18}. 

The tunability of suchlike Dirac-cone and flat-band physics is of particular importance from a technological perspective~\cite{Guea12}. It can be achieved applying external electric and magnetic fields, e.g., by nanoscale top gates, which modify the spatial electronic structure and allow to imprint junctions or barriers.  In this way the transport properties of gated graphene or Dice nanostructures can be manipulated.  In line with this and based on the parallels between optics and Dirac electronics,  the transmission of Dirac waves through potential barriers was intensively studied in the past, also beyond the Kein tunneling phenomenon~\cite{CPP07,BTB09,AU13,HBF13a,SHF15a,XL16}. 

For static quantum dots, different elastic scattering regimes have been identified~\cite{WF14}, ranging from quantum to quasiclassical behavior. Thereby, depending on the size of the dot and its energy to barrier height ratio, the barrier behaves as a resonant scatterer, a strong and weak reflector,  or a weak scatterer. Quite recently even Veselago lensing has been observed~\cite{Brea19}.  Equally interesting, close to resonances, long-living temporary bound states appear,  in spite of Klein tunneling~\cite{HA08}. Most of these theoretical results obtained within a continuum approach were confirmed for tight-binding lattice models by means of exact diagonalization techniques in the time following~\cite{UBWH11,PAS11,PHF13,PHWF14}.  Arrangements of ordered graphene nanodots also received  much attention~\cite{VAW11,Caea17}. In this field several theoretical predictions, such as a  Mie-type scattering of Dirac-Weyl particles~\cite{HBF13a} by quantum dot arrays~\cite{FHP15}, could be verified experimentally~\cite{CCOWK16}.

The problem of inelastic scattering of massless Dirac particles by an irradiated region on the other hand has received much less attention, probably because 
the associated Floquet scattering is much more difficult to treat. So far oscillating quantum dots were only studied for pseudo-spin-1/2 particles on the honeycomb lattice~\cite{SHF15b}. Compared to static quantum dots this leads to interesting observations even for weak (time-dependent) barrier modulations, such as a significant sideband-scattering  when the energies are located in the vicinity of avoided crossings of the (Floquet) quasienergy bands~\cite{WF18}. In this case interference effects cause a remarkably mixing of quantum and quasiclassical scattering behavior, which might have particularly interesting applications in the field of optomechanics, e.g., light-sound interconversion~\cite{WF17}.  Motivated by these results, we examine here the (even more involved) inelastic scattering of pseudo-spin-1 particles by a time-varying cylindrical barrier on the Dice lattice with a focus on the resonant scattering regime. 

The paper is organized as follows. In Sect.~\ref{section2}, we formulate and solve the corresponding scattering problem within Dirac-Weyl theory. Thereby we provide explicit expressions for the transmission  and reflection coefficients, as well as for the time-, distance- and angle-dependencies of the probability density, current density and scattering efficiency in the near- and far-field regions.  Furthermore, we make contact with important limiting cases.  All this is visualized and discussed in Sect.~\ref{section3}, which presents our main numerical results. There we also demonstrate the temporal trapping of the particle wave in vortex structures and its subsequent reemission with complex angle- and time-dependent radiation patterns. Most notably, for certain model parameters, we observe a forward and backward scattering that alternates in time. Also the revival of resonant scattering is observed in a broader energy region when compared with the static quantum dot. Our conclusions can be found in Sect.~\ref{section4}. 
 
 \begin{figure}[t]
\centering
\includegraphics[width=0.4 \textwidth]{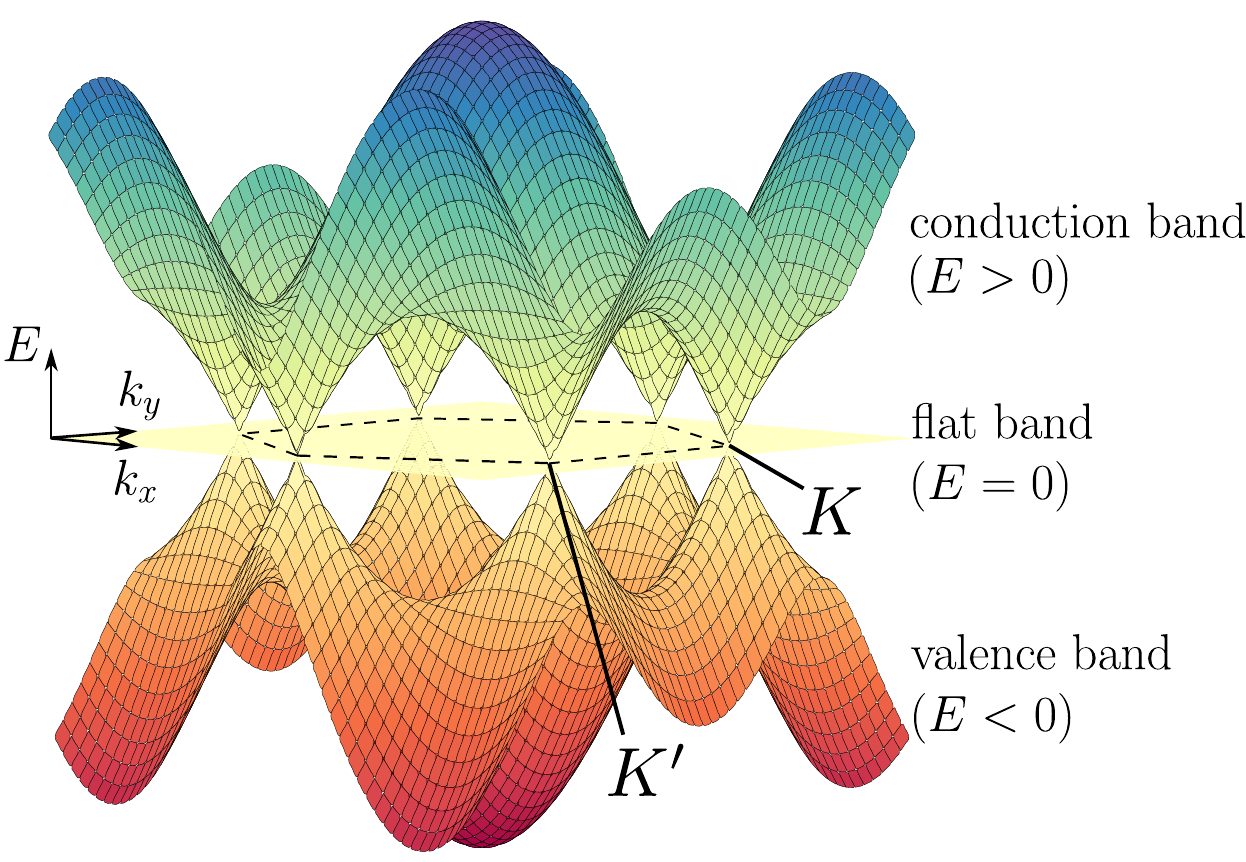}
\caption{Band structure of the tight-binding Dice-lattice model, containing in addition to the graphene valence and conduction bands a dispersionless (flat) band at $E=0$.}
\label{fig:dice_bandstructure}
\end{figure}
\begin{figure}[t]

\includegraphics[width=0.48 \textwidth]{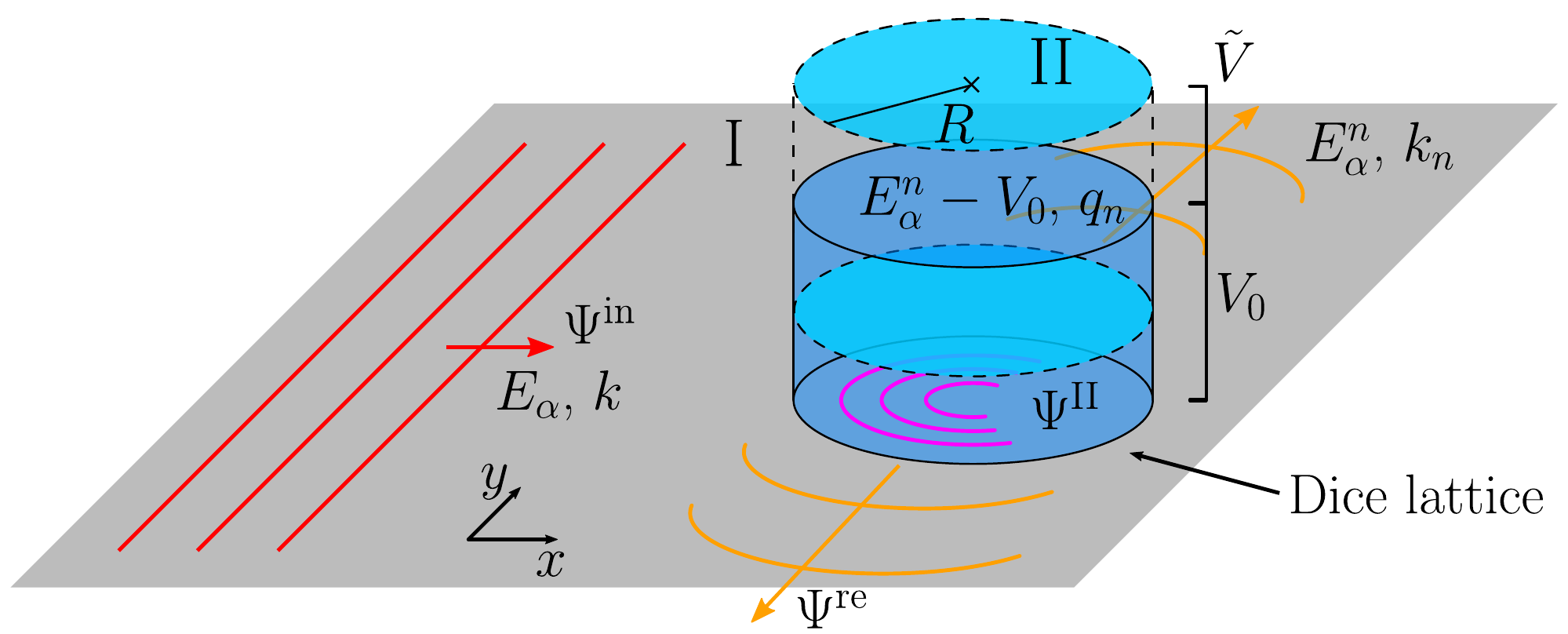}
\caption{Scattering setup considered in this work: A planar pseudospin-1 Dirac-Weyl wave with energy $E_\alpha$ and wave vector $k$, $\Psi^\text{in}$, propagating in positive $x$ direction on a 2D Dice-lattice, hits an oscillating quantum dot of radius $R$, which is realized by a gate-defined potential $V(r,t) = V_0 + \tilde{V}\cos{(\omega t)}$. I (II) marks the pure Dice lattice (gated region). As a result, reflected ($\Psi^\text{re}$) and transmitted waves ($\Psi^\text{tr}$) appear, with energies (momenta) $E_\alpha^n$ ($k_n)$ and $E_\alpha^n-V_0$ ($q_n$), respectively.  Here, $n$ denotes the sideband index.  }
\label{fig:osc_dot}
\end{figure}

 \section{Theoretical approach}
 \label{section2}
\subsection{Pseudospin-1 Dirac-Weyl model}
The perhaps most  simple model, taking into account the band structure of the Dice---or the more general $\alpha-\mathcal{T}_3$---lattice in the low-energy region near the corners of the Brillouin zone, seems to be the 2D generalized Dirac-Weyl Hamiltonian for massless spin-1 quasiparticles~\cite{BUGH09,Su86,UBWH11,VOVSDCD13,BCGC17,IN17},    
\begin{align}
H_\tau= v_\mathrm{F} \mathbf{S}_\tau \cdot \mathbf{p}, \label{eq:Dice-H}
\end{align}
which features a pair of Dirac cones intersected by a topologically flat band at the tip-touching points $K$ and $K'$, see Fig.~\ref{fig:dice_bandstructure}.
Here, $v_\mathrm{F}=3 at/\sqrt{2}\hbar$ denotes the Fermi (group) velocity, where  $a$  is the lattice constant and  $t$  is the nearest-neighbor transfer amplitude in a tight-binding model description, and $\mathbf{p}=-i\hbar \bm{\nabla}$ refers to  the 2D momentum operator. In what follows we set $\hbar=1$. The pseudospin vector $\mathbf{S}_\tau=(\tau S_x, S_y)$, with
\begin{align}
S_x &= \frac{1}{\sqrt{2}}\begin{pmatrix}
0 & 1 & 0 \\
1 & 0 & 1 \\
0 & 1 & 0 
\end{pmatrix},
& S_y &= \frac{1}{\sqrt{2}} \begin{pmatrix}
0 & -i & 0 \\
i & 0 & -i \\
0 & i & 0 
\end{pmatrix} ,
\label{eq:spin1xy-matrices}
\end{align}
represents the sublattice degree of freedom, i.e., the valley index  $\tau=+1$ $(-1)$ for the $K$ ($K'$) point. Note that the pseudospin-1 matrices fulfil the commutator relations of a angular momentum algebra $[S_i,S_j]=i \varepsilon_{ijk} S_k$ with
\begin{align}
& S_z = \begin{pmatrix}
1 & 0 & 0 \\
0 & 0 & 0 \\
0 & 0 &-1
\end{pmatrix},
\label{eq:spin1z-matrices}
\end{align}
but not those of the Clifford algebra ($\{S_i,S_j\}= \delta_{ij} \mathbb{I}_{3}$) being valid for  the graphene pseudospin-1/2 system.

Let us now consider a cylindrical, harmonically driven potential barrier   
\begin{align}
V(r,t) = \left(V_0 + \tilde{V}\cos{(\omega t)}\right)\Theta(r-R)\,,
\label{eq:potential}
\end{align}
which, in a way, realizes a gate-defined quantum dot of radius $R$, see Fig.~\ref{fig:osc_dot}. In Eq.~\eqref{eq:potential}, $V_0$ embodies a static barrier  and 
$\tilde{V}$  denotes the amplitude of the potential part that oscillates in time with angular frequency  $\omega$. Both $V_0$ and $\tilde{V}$ are assumed to vanish outside the gated region. The use of such a step-like potential in conjunction with the single-valley continuum approximation (neglecting  any intervalley scattering)  is a good approximation for sufficiently low energies, and barrier potentials  that are smooth on the scale of the lattice constant but sharp on the scale of the de Broglie wave length. 
The validity of this has been proven at least for pseudospin 1/2 case by comparison with the exact numerical solution of the full (tight-binding model based) scattering problem~\cite{PHF13}.  As a result, the effective Hamiltonian in the vicinity of the $K$ (or $K'$) point is  
\begin{align}
H(\mathbf{r},t) = v_{\mathrm{F}} \mathbf{S} \cdot \mathbf{p}+ V(r,t) \mathbb{I}_{3}\,, \label{eq:ges_H}
\end{align}
i.e., we can suppress the index $\tau$ hereafter. 
\subsection{Solution of the scattering problem}
Treating the inelastic scattering problem (due the oscillating barrier the quasiparticle may exchange energy quanta $n \omega$ with the external field), we look for solutions $\Psi(\mathbf{r}, t)$ of the time-dependent Dirac-Weyl equation $i(\partial/\partial t)\Psi({\mathbf{r},t})=H(\mathbf{r},t) \Psi(\mathbf{r},t)$ in the complete plane. Thereby, with a view to the scattering geometry, we expand the wave functions of the incident plane wave $\Psi^{\text{in}}$ (propagating in $x$ direction), the reflected (scattered) wave $\Psi^{\text{re}}$, and the transmitted wave  $\Psi^{\text{tr}}$ in polar coordinates $r$ and $\phi$.  
In region I ($r>R$, see Fig.~\ref{fig:osc_dot}), the wave function  $\Psi^{\text{I}}(r,\phi,t)= \Psi^{\text{in}}(r,\phi,t)+\Psi^{\text{re}}(r,\phi,t)$  is build with 
\begin{align}
\Psi^{\text{in}}(r,\phi,t)=  \sum \limits_{n;m=-\infty}^{\infty}\delta_{n,0}\,i^{m-1} \psi_{m, E_\alpha^n}^{(0)}(r,\phi) e^{-i E^n_\alpha t}\,, \label{eq:wvfkt_i_inc}\\
\Psi^{\text{re}}(r,\phi,t)=\sum \limits_{n;m=-\infty}^{\infty} r_{mn} \,i^{m-1}\psi^{(1)}_{m,E_\alpha^n}(r,\phi) e^{-i E^n_\alpha t},  \label{eq:wvfkt_i_refl}
\end{align}
where \begin{align}
\psi_{m,E_\alpha^n}^{(0,1)} =\frac{1}{2} \begin{pmatrix}
Z_{m-1}^{(0,1)}(k_n r) e^{i(m-1)\phi} \\
i \alpha_n \sqrt{2} Z_{m}^{(0,1)}(k_n r) e^{im\phi} \\
-Z_{m+1}^{(0,1)}(k_n r) e^{i(m+1)\phi}
\end{pmatrix}\;, \label{eq:wvf-freeparticle-polar}
\end{align}
are the eigenfunctions for the dispersive bands 
\begin{align}
E_\alpha^n=E_\alpha +n \omega\,,
\label{eq:esn}
\end{align}
and
\begin{align}
\psi_{m,0}^{(0,1)}=\frac{1}{2} \begin{pmatrix}
Z_{m-1}^{(0,1)}(k r) e^{i(m-1)\phi} \\
0 \\
Z_{m+1}^{(0,1)}(k r) e^{i(m+1)\phi}
\end{pmatrix}
\end{align}
or those for the flat $E=0$ band. 
Here, $E^n_\alpha=\alpha_n v_\mathrm{F}k_n$ (with $\alpha_n=\text{sgn} (E_\alpha^n)$ and $n \in \mathbb{Z}$), $E_\alpha=\alpha v_{\mathrm{F}} k$, $\alpha=\text{sgn} (E_\alpha)$, and $k=|\mathbf{k}|$. Furthermore, $Z_m^{(0)}=J_m$ is the Bessel function of first kind. $Z_m^{(1)}=H_m$ denotes the Hankel function of first or second kind, depending on the energy sideband index $n$ in $H_m(k_n r)=J_m(k_nr)+\alpha_n Y_m(k_n r)$ where $Y_m(k_n r)$ is the Bessel function of second kind (Neumann function). In Eqs.~\eqref{eq:wvfkt_i_inc} and~\eqref{eq:wvfkt_i_refl}, the summations  are performed over $n$ and  the angular momentum quantum number $m$; the $r_{mn}$ are the scattering coefficients of the reflected wave.   

In region II ($r<R$), the potential $V(r,t)$ causes an explicit time dependence of the Hamiltonian~\eqref{eq:ges_H}. Inserting the separation ansatz, 
 $\Psi^{\text{II}}(\mathbf{r},t)=\Psi^{\text{tr}}(\mathbf{r},t)=\Psi(\mathbf{r}) \xi(t)$, we find
\begin{align}
i \frac{\dot{\xi}(t)}{\xi(t)}-\tilde{V}\cos(\omega t) = v_\mathrm{F}\frac{\mathbf{S}\cdot \mathbf{p}\; \Psi(\mathbf{r})}{\Psi(\mathbf{r})} + V_0\,,
\end{align}
yielding
\begin{align}
i \frac{\dot{\xi}(t)}{\xi(t)}-\tilde{V}\cos(\omega t) = c \label{eq:sep-Var.-Zeit}\,,\\[0.2cm]
v_\mathrm{F}\frac{\mathbf{S}\cdot \mathbf{p} \,\Psi(\mathbf{r})}{\Psi(\mathbf{r})} + V_0 = c\,. \label{eq:sep-Var.-Ort}
\end{align}
Identifying $c=E_\alpha$, Eq.~\eqref{eq:sep-Var.-Ort} turns out to be the stationary pseudospin-1 Dirac-Weyl equation. Integration of Eq.~\eqref{eq:sep-Var.-Zeit}  yields
\begin{align}
\xi(t) = c_1 e^{-iE_\alpha t} \cdot e^{-i\frac{\tilde{V}}{\omega}\sin(\omega t)}\,.
\end{align}
Overall, we obtain in region II 
\begin{align}
\Psi^{\text{II}}(r,\phi,t) =  \sum \limits_{n;m=-\infty}^{\infty}& t_{mn}\,i^{m-1} \psi_{m,E_\alpha^n-V_0}^{(0)}(r,\phi)\nonumber\\&\times  e^{-iE_\alpha^n t} e^{-i\frac{\tilde{V}}{\omega}\sin{(\omega t )}}\label{eq:wfkt_ii_1}
\end{align}
with transmission amplitudes $t_{mn}$ and $ E_\alpha^n-V_0 = \bar{\alpha}_n v_\mathrm{F}q_n $. By means of the Jacobi-Anger identity, Eq.~\eqref{eq:wfkt_ii_1} can be brought into the form
\begin{align}
\Psi^{\text{II}}(r,\phi,t) = \sum \limits_{n,p;m=-\infty}^{\infty}&t_{mn}\,i^{m-1} \psi_{m,E_\alpha^n-V_0}^{(0)}(r,\phi)\nonumber\\&\times  e^{-i(E_\alpha^n-p \omega) t}(-1)^p J_p\left(\frac{\tilde{V}}{\omega}\right)\,. \label{eq:wfkt_ii}
\end{align} 
If not otherwise specified, all summations will run from minus to plus infinity below.

\subsection{Reflection and transmission coefficients}
We now determine the scattering coefficients  $r_{mn}$ and $t_{mn}$ by matching the three components $\psi_j$ ($j=1,2,3$)  of the Dirac-Weyl spinor $\Psi$ at $r=R$ ~\cite{UBWH11}:
\begin{align}
\psi_2^{\text{I}}(R) &= \psi_2^{\text{II}}(R)\,, \\
e^{i\phi}\psi_1^{\text{I}}(R) + e^{-i\phi}\psi_3^{\text{II}}(R) &= e^{i\phi}\psi_1^{\text{II}}(R) + e^{-i\phi}\psi_3^{\text{II}}(R)\,. \label{eq:wvf-continuity-dot}
\end{align}
Then, for the current operator $\mathbf{j}=v_{\text{F}}( S_x, S_y)$, the radial component is continuous at the boundary
\begin{align}
\Psi^\dagger \mathbf{e}_r \cdot \mathbf{j} \Psi = \sqrt{2}v_\mathrm{F} \Re \left[\psi_2^\ast(\psi_1 e^{i\phi} + \psi_3 e^{-i\phi})\right]\,, \label{eq:continuity-radial}
\end{align}  
but not necessarily the tangential component
\begin{align}
\Psi^\dagger\mathbf{e}_\phi \cdot \mathbf{j} \Psi = -\sqrt{2} v_\mathrm{F} \Im \left[\psi_2^\ast(\psi_1 e^{i\phi} - \psi_3 e^{-i\phi}) \right]\,. \label{eq:continuity-azimutal}
\end{align}

Inserting the spinor wave function, the time-dependent phase factors agree if the quantized wave numbers are $v_\mathrm{F}k_n=\alpha_n E_\alpha^n$ and $v_\mathrm{F} q_p=\bar{\alpha}_p (E_\alpha^p -V_0)$  in regions I and II, respectively. Then, utilizing   $\mathcal{Z}^{(0)}_m(x)= J_{m-1}(x)-J_{m+1}(x)$ and $\mathcal{Z}^{(1)}_m(x)=H_{m-1}(x)-H_{m+1}(x)$,  the conditional equations for $r_{mn}$ and $t_{mn}$ become: 
\begin{align}
\delta_{n,0} &\mathcal{Z}^{(0)}_m(k_n R) +r_{mn} \mathcal{Z}^{(1)}_m(k_n R)\nonumber\\ &= \sum \limits_p t_{mp} \mathcal{Z}^{(0)}_m(q_p R)  (-1)^{n-p} J_{n-p}\left(\frac{\tilde{V}}{\omega}\right) \label{eq:stetigkeit-oszillierend1} \,,\\
\delta_{n,0} &\alpha_n J_m(k_n R) +r_{mn} \alpha_n H_m(k_n R)\nonumber\\ & = \sum \limits_p t_{mp} \bar{\alpha}_p J_m(q_p R) (-1)^{n-p}J_{n-p}\left(\frac{\tilde{V}}{\omega}\right)\,.\label{eq:stetigkeit-oszillierend2}
\end{align}
Now, multiplying Eq.~\eqref{eq:stetigkeit-oszillierend1}  by  $\alpha_n H_m(k_n R)$, Eq.~\eqref{eq:stetigkeit-oszillierend2} by $\mathcal{Z}^{(1)}_m(k_n R)$, and subtracting the resulting equations, we arrive at 
\begin{align}
\delta_{n,0}g_m^{(n)} &=\sum \limits_p t_{mp} (-1)^{n-p} J_{n-p}\left(\frac{\tilde{V}}{\omega}\right) f_m^{(n,p)} \label{eq:r_mn-components}
\end{align}
with the substitutions
\begin{align}
f_m^{(n,p)}& = \mathcal{Z}^{(0)}_m(q_p R)H_m(k_n R) - \alpha_n \bar{\alpha}_p J_m(q_p R)\mathcal{Z}^{(1)}_m(k_n R), \\
g_m^{(n)}&= \mathcal{Z}^{(0)}_m(k_n R)H_m(k_n R) -J_m(k_n R) \mathcal{Z}^{(1)}_m(k_n R)\, .
\end{align}
That means  the reflection coefficients result from 
\begin{align}
r_{mn} = \sum \limits_p &t_{mp} \bar{\alpha}_p (-1)^{n-p} J_{n-p}\left(\frac{\tilde{V}}{\omega}\right) \frac{J_m(q_p R)}{\alpha_n H_m(k_n R)}\nonumber\\
&-\delta_{n,0}\frac{J_m(k_n R)}{H_m(k_n R)}. \label{eq:r_mn}
\end{align}

Calculating the transmission coefficients $t_{mn}$, we rewrite Eq.~\eqref{eq:r_mn-components} with $J_{n-p}(\tilde{V}/\omega)\equiv J_{n-p}$ as an infinite set of equations,
\begin{align}
\mathbf{M}_m \mathbf{t}_m = \mathbf{g}_m, \label{eq:r_mn-matrix}
\end{align}
with the vectors $\mathbf{t}_m = (\dots,\, t_{m,\,-1},\, t_{m,\,0},\,t_{m,\,+1},\,\dots)$,\\ ${\mathbf{g}_m = (\dots,\,0,\, g_m^{(0)},\,0,\, \dots)}$ and the matrix 
\begin{align}
\mathbf{M}_m = \begin{pmatrix}
&	& \vdots 	&  \\
& J_ 0 f_m^{(-1,-1)} & J_{-1} f_m^{(-1,0)} & J_{-2}f_m^{(-1,1)} & \\
\cdots &  J_{1} f_m^{(0,-1)} & J_0 f_m^{(0,0)} & J_{-1} f_m^{(0,1)} & \cdots\\
&  J_{2}f_m^{(1,-1)}  &  J_{1} f_m^{(1,0)} & J_0 f_m^{(1,1)}  & \\
&	& \vdots 	& 
\end{pmatrix}\,.
\end{align}
In the numerical solution of~\eqref{eq:r_mn-matrix}, we gradually increase the dimension of the system of equations until convergence is reached. Thereby, the ratio 
$\tilde{V}/\omega$, indicating how many sidebands $n$ will be relevant,  allows to estimate the matrix dimension, i.e., to formulate a truncation condition. 

For small $x=\tilde{V}/\omega$, the Bessel and von Neumann functions can be expanded in $x$: 
\begin{align}
J_m(x) &\simeq \begin{cases}\frac{\left(-1\right)^m}{(-m)!}\left(\frac{2}{x}\right)^m &\qquad\qquad  \text{if } m<0,\\
 1 &\qquad\qquad  \text{if } m=0, \\
   \frac{1}{m!}\left(\frac{x}{2}\right)^m &\qquad\qquad  \text{if } m>0,
\end{cases}\\[.4cm]
 Y_m(x) &\simeq \begin{cases}  
-\frac{(-1)^m (-m-1)!}{\pi}\left(\frac{x}{2}\right)^m  &\text{if } m<0, \\
\frac{2}{\pi} \left(\ln\frac{x}{2}+\gamma_{\rm E}\right) & \text{if } m=0, \\
-\frac{(m-1)!}{\pi}\left(\frac{2}{x}\right)^m & \text{if } m>0\,,
\end{cases} \label{eq:J-Y-series_expansion}
\end{align}
where $\gamma_{\rm E}$ is the Euler-Mascheroni constant. In the limit $x\to 0$ (static barrier), we have $J_0(0)=1$ and $J_{n\neq 0}(0)=0$, and the matrix $\mathbf{M}_m$ reduces to
\begin{align}
M_m = f_m^{(n,p)} \delta_{n,0} \delta_{p,0}\,,
\end{align}
i.e., $t_{mn}=r_{mn}=0$ for any finite $n$. Then we get
\begin{align} 
r_{m,0} &= -\frac{J_m(q_0R)\mathcal{Z}^{(0)}_m(kR_0)-\alpha\bar{\alpha}\mathcal{Z}^{(0)}_m(q_0R)J_m(kR)}{J_m(q_0R)\mathcal{Z}^{(1)}_m(k_0R)-\alpha\bar{\alpha}\mathcal{Z}^{(0)}_m(q_0R)H_m(k_0R)}, \\
t_{m,0} &= \frac{H_m(k_0R)\mathcal{Z}^{(0)}_m(k_0R)-\mathcal{Z}^{(1)}_m(k_0R) J_m(k_0R)}{H_m(k_0R)\mathcal{Z}^{(0)}_m(q_0R)-\alpha\bar{\alpha}\mathcal{Z}^{(1)}_m(k_0R)J_m(q_0R)}
\end{align}
with  $\bar{\alpha}=\bar{\alpha}_{n=0}$.  
\subsection{Scattering characteristics}
\subsubsection{Near-field}
In this section we specify the expressions for the time-dependent probability density $\rho=\langle\Psi|\Psi\rangle$ and probability current density   $\mathbf{j} = \langle \Psi| v_\mathrm{F} \mathbf{S}|\Psi \rangle$.  

In region I $(r>R)$, we have $\rho^{\text{I}} = \rho^{\text{in}} + \rho^{\text{re}} + \rho^{\text{if}}$ and $j_{x,y}^{\text{I}} = j_{x,y}^{\text{in}} + j_{x,y}^{\text{re}} + j_{x,y}^{\text{if}}$ with $\rho^{\text{in}}=1$,  $ j_{x}^{\text{in}}=v_ \mathrm{F}\alpha $  and $ j_{y}^{\text{in}}=0$. The reflected contributions have to be calculated from $\rho^{\text{re}}=\langle\Psi^{\text{re}}|\Psi^{\text{re}}\rangle$ and $j_{x,y}^{\text{re}} = v_\mathrm{F} \langle\Psi^{\text{re}}|S_{x,y}| \Psi^{\text{re}}\rangle$. The interference contributions are obtained according to  $\rho^{\text{if}}=2\Re\langle\Psi^{\text{in}}|\Psi^{\text{re}}\rangle $  and $ j_{x,y}^{\text{if}}=  2v_\mathrm{F} \Re \langle \Psi^{\text{in}}|S_{x,y}| \Psi^{\text{re}}\rangle$. Inserting the individual wave functions, we find
\begin{align}
\rho^{\text{re}}=& \frac{1}{4}\sum \limits_{\substack{m,n \\l,p}} r_{lp}^\ast r_{mn} \,i^{m-l} e^{i(m-l)\phi}e^{i(p-n)\omega t}\\
&\times\left[2 \alpha_p \alpha_n H_l^\ast(k_p r)H_m(k_n r)+H_{l-1}^\ast(k_p r)H_{m-1}(k_n r)\right. \nonumber\\&\qquad\left.+H_{l+1}^\ast(k_p r)H_{m+1}(k_n r)\right]\nonumber,
\\[0.4cm]
\rho^{\text{if}}=& \frac{1}{2}\Re \big\{e^{-ikr\cos\phi} \sum \limits_{\substack{m,n \\}} r_{mn}\,i^{m-1}e^{-in\omega t}e^{im\phi}\\& \times \left[2 i \alpha \alpha_n H_m(k_nr)+H_{m-1}(k_n r)e^{-i\phi}\right. \nonumber \\& \left.-H_{m+1}(k_nr)e^{i\phi} \right] \big\},\nonumber
\end{align}
and
\begin{align}
j_x^{\text{re}}=& \frac{v_\mathrm{F}}{4}\sum \limits_{\substack{m,n\\ l,p}}r_{lp}^\ast r_{mn} \, i^{m-l+1} e^{i(m-l)\phi}  e^{i(p-n)\omega t}\\&\times 
 \left[ \alpha_n H_m(k_n r)\left( H_{l-1}^\ast(k_p r) e^{i\phi}- H_{l+1}^\ast(k_p r) e^{-i\phi}\right) \right.\nonumber\\
&\left. -\alpha_p H_l^\ast(k_p r) \left( H_{m-1}(k_n r) e^{-i\phi}-H_{m+1}(k_n r) e^{i\phi}\right)\right],\nonumber\\[0.4cm]
j_x^{\text{if}}=& \frac{v_\mathrm{F}}{2} \Re\big\{ e^{-ikr\cos\phi}\sum \limits_{m,n}r_{mn}\,i^{m-1} e^{-in\omega t}  e^{i m \phi}  \\
 &\times \left[i2\alpha_n H_m(k_n r) +\alpha\left( H_{m-1}(k_n r)e^{-i\phi} \right.\right.\nonumber\\&\qquad\left.\left.- H_{m+1}(k_n r) e^{i\phi}\right) \right]\big\},
\nonumber\\[0.4cm]
j_y^{\text{re}} = &\frac{v_\mathrm{F}}{4}\sum \limits_{\substack{m,n\\ l,p}}r_{lp}^\ast r_{mn} \,i^{m-l} e^{i(m-l)\phi} e^{i(p-n)\omega t}  \\
 &\times \left[ \alpha_n H_m(k_n r)\left( H_{l-1}^\ast(k_p r) e^{i\phi}+ H_{l+1}^\ast(k_p r) e^{-i\phi}\right) \right.\nonumber\\
&\left. +\alpha_p H_l^\ast(k_p r) \left( H_{m-1}(k_n r) e^{-i\phi}+ H_{m+1}(k_n r) e^{i\phi}\right)\right]\nonumber,\\[0.4cm]
j_y^{\text{if} }=&  \frac{v_\mathrm{F}}{2} \Re \big\{e^{-ikr\cos\phi} \sum \limits_{m,n}r_{mn}\,i^{m} e^{-in\omega t} e^{i m \phi}   \\
 &\times \left[\alpha\left( H_{m-1}(k_n r)e^{-i\phi} + H_{m+1}(k_n r) e^{i\phi}\right) \right]\big\}.\nonumber
\end{align}

In region II $(r<R)$, we have $\rho^{\text{II}} = \rho^{\text{tr}}=\langle\Psi^{\text{tr}}| \Psi^{\text{tr}}\rangle$ and $j_{x,y}^{\text{II}} = j_{x,y}^{\text{tr}}= v_\mathrm{F} \langle\Psi^{\text{tr}}|S_{x,y} |\Psi^{\text{tr}}\rangle$ with
 \begin{align}
\rho^{\text{tr}}=& \frac{1}{4}\sum \limits_{\substack{m,n \\l,p}}t_{lp}^\ast t_{mn}  \, i^{m-l} e^{i(m-l)\phi}e^{i(p-n)\omega t} \\
&\times \left[2 \bar{\alpha}_p \bar{\alpha}_n J_l(q_pr)J_m(q_nr)+J_{l-1}(q_pr)J_{m-1}(q_n r)\right.\nonumber\\&\left.\qquad+J_{l+1}(q_pr)J_{m+1}(q_n r)\right] \nonumber
\end{align}
and
\begin{align}
j_x^{\text{tr}}=& \frac{v_\mathrm{F}}{4}\sum \limits_{\substack{m,n\\ l,p}} t_{lp}^\ast t_{mn} \,i^{m-l+1}e^{i(m-l)\phi} e^{i(p-n)\omega t} \\
& \left[ \bar{\alpha}_n J_m(q_n r)\left( J_{l-1}(q_p r) e^{i\phi}-J_{l+1}(q_p r) e^{-i\phi}\right) \right.\nonumber\\
&\left. -\bar{\alpha}_n J_l(q_p r) \left( J_{m-1}(q_n r) e^{-i\phi}-J_{m+1}(q_n r) e^{i\phi}\right)\right],\nonumber\\[0.4cm]
j_y^{\text{tr}} =& \frac{v_\mathrm{F}}{4}\sum \limits_{\substack{m,n\\ l,p}}t_{lp}^\ast t_{mn} \,i^{m-l} e^{i(m-l)\phi} e^{i(p-n)\omega t} \\
& \left[ \bar{\alpha}_n J_m(q_n r)\left( J_{l-1}(q_p r) e^{i\phi}+ J_{l+1}(q_p r) e^{-i\phi}\right) \right.\nonumber\\
&\left. +\bar{\alpha}_p J_l(q_p r) \left( J_{m-1}(q_n r) e^{-i\phi}+J_{m+1}(q_n r) e^{i\phi}\right)\right],\nonumber
\end{align}
respectively.
\subsubsection{Far-field}
We next consider the time-dependent, radial, reflected current density in the far-field, $j_{r}^{\text{re}} = v_\mathrm{F}\langle\Psi^{\mathrm{re}}|\mathbf{S} \cdot\mathbf{e}_r |\Psi^{\mathrm{re}}\rangle$, where
\begin{align}
\mathbf{S} \cdot \mathbf{e}_r  =\frac{1}{\sqrt{2}}\begin{pmatrix}
0 & \, e^{-i\phi} & 0 \\
 e^{i\phi} & 0 & e^{-i\phi} \\
0 & e^{i\phi} & 0 
\end{pmatrix}.
\end{align}

To leading order in $1/r$ we obtain after a lengthy but straightforward calculation: \\
\begin{align}
j_r^{\text{re}}(r,\phi,t)=& \frac{2v_\mathrm{F}}{\pi r}\Bigg\{\sum \limits_n \bigg[\sum \limits_m \frac{|r_{mn}|^2}{k_n}\label{eq:full_ff_curr} \\&+ \Re {\sum \limits_{\substack{m>l\\ m\neq l}} \frac{r_{mn} r_{ln}^\ast}{k_n}} e^{i\frac{\pi}{2}(1-\alpha_n)(m-l)} e^{i(m-l)\phi}   \bigg] \nonumber\\
&+ \Re \bigg[\sum \limits_{\substack{n>p\\n\neq p}}\frac{(1-i\alpha_n)(1+i\alpha_p)}{\sqrt{k_n k_p}} e^{i(n-p)\omega(r/v_\mathrm{F}-t)} \nonumber\\
&\qquad\times \bigg( \sum \limits_m r_{mn} r_{mp}^\ast e^{i\frac{\pi}{2}(m-1)(\alpha_p-\alpha_n)} \nonumber\\&\qquad\quad+  2\sum \limits_{\substack{m>l\\ m\neq l}} r_{mn} r_{lp}^\ast e^{i\frac{\pi}{2}(m-1)(1-\alpha_n)} \nonumber\\&\qquad\quad\times e^{-i\frac{\pi}{2}(l-1)(1-\alpha_p)} e^{i(m-l)\phi} \bigg)\bigg]\Bigg\}\nonumber\,.
\end{align}
Using $\frac{1}{T}\int \limits_t^{t+T} e^{i(n-p)\omega t'} \mathrm{d}t' = \delta_{np}$, we find for the time average of the radial reflected current:
\begin{align}
\overline{j_r^{\text{re}}}(r,\phi) =& \frac{2v_\mathrm{F}}{\pi r}\sum \limits_n \bigg[\sum \limits_m \frac{|r_{mn}|^2}{k_n} \\&+ \Re {\sum \limits_{\substack{m>l\\ m\neq l}} \frac{r_{mn} r_{ln}^\ast}{k_n}  e^{i\frac{\pi}{2}(1-\alpha_n)(m-l)} e^{i(m-l)\phi} } \bigg] . \nonumber\label{eq:ff_curr_ta}
\end{align}
\subsubsection{Scattering efficiency }
The scattering of a Dirac-Weyl quasiparticle on a circular potential step is advantageously discussed in terms of the scattering efficiency, i.e.,  the scattering cross section divided by the geometric cross section~\cite{HBF13a}, 
\begin{align}
Q=\frac{\sigma}{2R}\,,
\end{align}
where $\sigma=I^\text{re}/\alpha v_\text{F}$ is the  quotient between the reflected probability current $I^\text{re}=\int_0^{2\pi}j_r^\text{re} r \text{d}\phi$ and the  total incoming current per unit area $(I^\text{in}/A)=v_\text{F}\alpha$. Inserting the reflected radial current in far-field approximation~\eqref{eq:full_ff_curr}, we get for the time-dependent scattering efficiency
\begin{align}
Q =& \frac{2}{R} \sum \limits_m \bigg[\sum \limits_n \frac{|r_{mn}|^2}{k_n} + \Re \sum \limits_{\substack{n>p\\n\neq p}}\frac{(1-i\alpha_n)(1+i\alpha_p)}{\sqrt{k_n k_p}}\nonumber\\&\quad\times r_{mn} r_{mp}^\ast e^{i\frac{\pi}{2}(m-1)(\alpha_p-\alpha_n)}e^{i(n-p)\omega\big(\tfrac{r}{v_\mathrm{F}}-t\big)}\bigg] \label{eq:Q-time}
\end{align}
and, after time averaging,
\begin{align}
\overline{Q} = \frac{2}{R}\sum \limits_{m,n} \frac{|r_{mn}|^2}{k_n}\,. \label{eq:Q-osc}
\end{align}
Note that Eq.~\eqref{eq:Q-osc} reduces to the result for the static quantum dot if one ignores the summation over the sidebands $n$. 

\section{Numerical results and discussion}
 \label{section3}
\begin{figure}[htb]
\includegraphics[width=0.45 \textwidth]{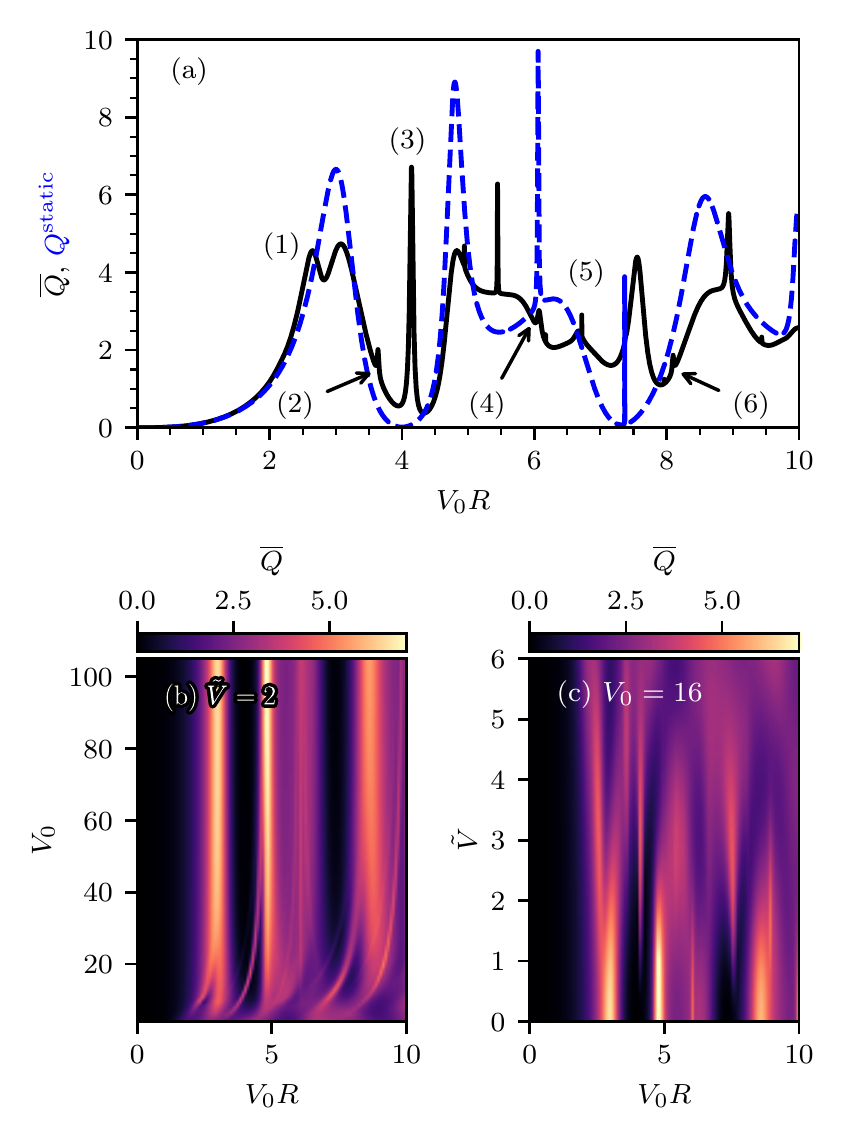}
\caption{Time-averaged scattering efficiency $\overline{Q}$ as a function of $V_0R$  for a  pseudospin-1 Dirac-Weyl wave hitting an oscillating quantum dot with energy to (static) barrier-height ratio $E/V_0=0.1$. Top panel (a):  $\overline{Q}$ for $V_0=16$ and $\tilde{V}=2$ (black line) compared to the static case ($\tilde{V}=0$; blue dashed line). Here, for the subsequent discussion, selected resonances are marked by numbers. Bottom panels: Intensity plot of $\overline{Q}$ 
varying $V_0$ at fixed $\tilde{V}=2$ (b) respectively $\tilde{V}$ at fixed $V_0=16$ (c).}
\label{fig:fig3}
\end{figure}

\begin{figure}[htb]
\includegraphics[width=0.45 \textwidth]{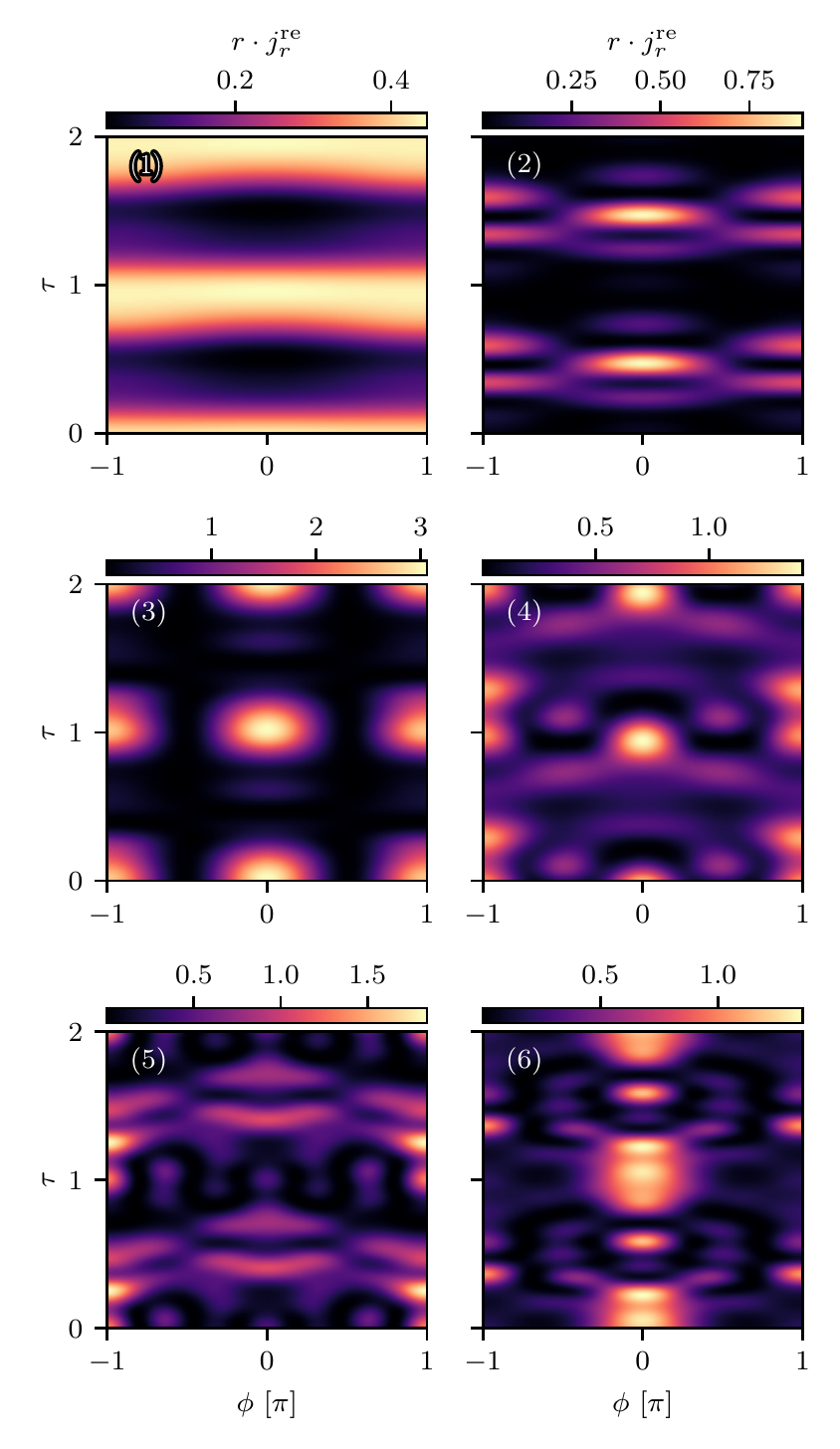}
\caption{Intensity plot of current density of the reflected wave in the far-field, $r\cdot j_r^{\text{re}}(r,\phi,t)$ from Eq.~\eqref{eq:full_ff_curr}. 
Results are given in the scattering-angle ($\phi$) -- time ($\tau$) domain at the resonances (parameters) indicated by (1),$\ldots$,(6) in Fig.~\ref{fig:fig3}.} 
\label{fig:fig4}
\end{figure}

\begin{figure}[htb]
\includegraphics[width=0.45 \textwidth]{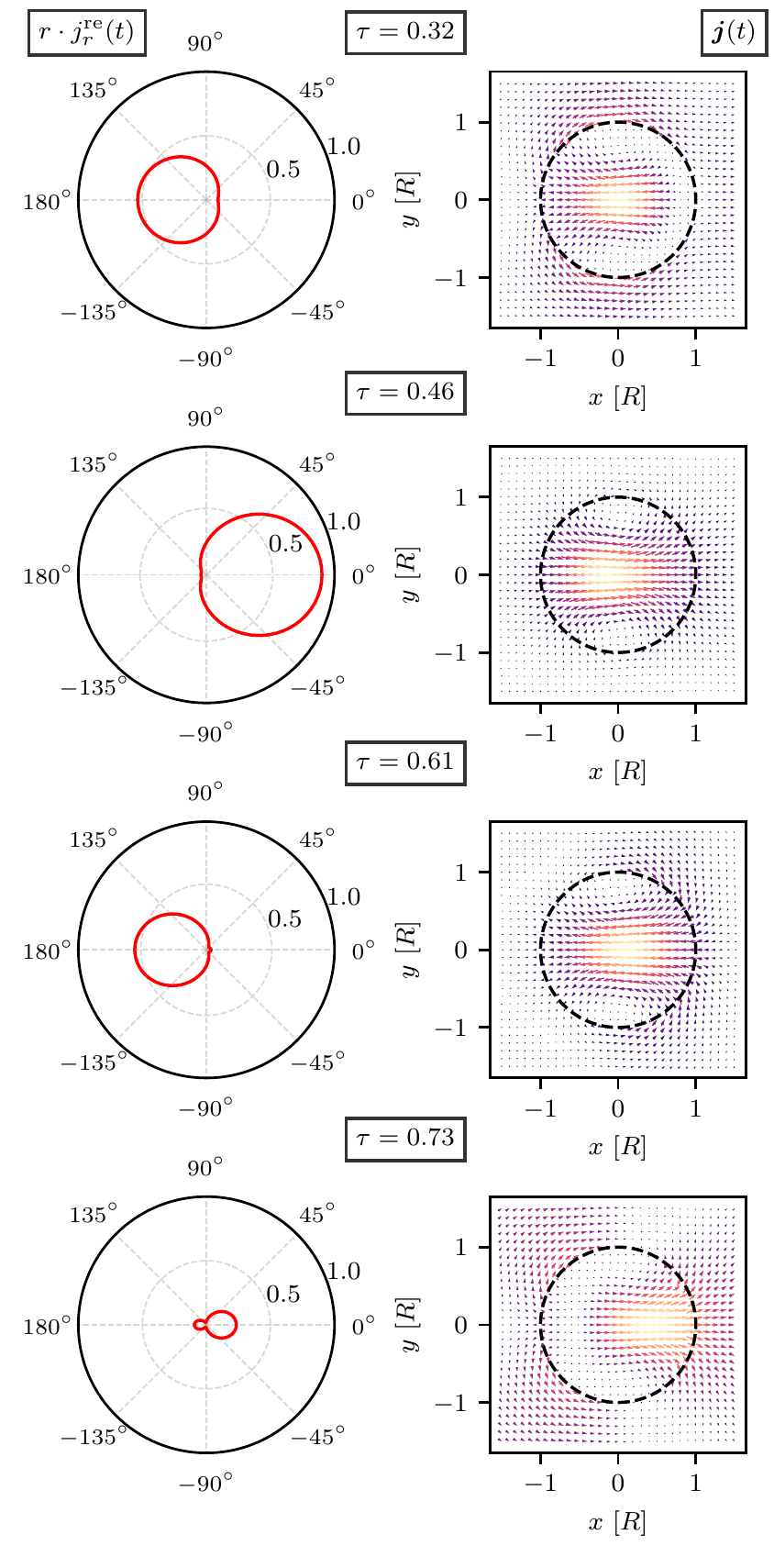}
\caption{Left panels: Polar plot of the reflected current density in the far-field at various time steps. Right panels: Current density in the near-field at the same time steps. Here, the arrow lengths (colors) are given relative to its maximum absolute values, which are  5.376, 26.499,  20.824, and 
4.015 (from top to bottom). Results are for $E/V_0=0.1$, $V_0=16$, $\tilde{V}=2$ at $V_0R=3.6373$ [resonance (2)].}
\label{fig:fig5}
\end{figure}

The scattering of plane Dirac waves on a {\it static} cylindrical, electrostatically realized potential barrier (quantum dot) was analyzed for both graphene (pseudospin-1/2)~\cite{CPP07,HBF13a,AU13,AU14,SHF15a,SHF15b,WF17} and Dice~\cite{XL16}  (pseudospin-1) quasiparticles. It was found that different scattering regimes are realized depending on two parameters, $ER$  and $V_0R$, which specify the size and the strength of  the barrier, respectively. Hereafter, we use units such that  $v_{\rm F}=1$) and limit ourselves to $\alpha>0$, i.e., $E_n^\alpha=E_n$ without loss of generality. $ER$ also determines the maximum angular momentum being possible in the scattering~\cite{WF14}. At small (very large) size parameters, $ER\leq1$ ($ER\gg1$), the particle's wavelength is larger (much smaller) than the radius of the quantum dot and only a few (many) partial waves will significantly affect the scattering process, which means  the quantum (quasi-classical) regime is realized. In the so-called resonant scattering  regime, where $E/V_0\ll 1$ (cf. Fig.~2 in Ref.~\cite{WF18}), the excitation of the first partial waves causes sharp resonances in the scattering efficiency, which can be viewed as quasi-bound quantum dot states. Increasing $E/V_0$, more and more partial waves will be excited and the quantum dot acts as a strong reflector (or even as a wave forward focusing Veselago lens) as long as $E/V_0\lesssim1$,  as a weak reflector if $E/V_0>1$, and finally as a weak scatterer provided that $V_0R<1$. One special feature of the Dice quantum dot---if compared with the graphene-based one---is the revival resonant scattering phenomenon that  occurs at about $E/V_0=0.5$. It occurs because of a peculiar boundary trapping profile  originating from fusiform vortices~\cite{XL16}.  Note that this revival resonant scattering is quite robust and persists even in the fairly short-wavelength regime~\cite{XL19}.  

In what follows, we will mainly focus on the resonant scattering regime ($E/V_0\ll 1$). Here,  in the static case,  Eq.~\eqref{eq:J-Y-series_expansion} gives the resonance condition 
\begin{align}
V_0R &\simeq \begin{cases}  
j_{0}^{(p)}-ER \ln (e^{\gamma_{\rm E}}ER/2) & \text{if } m=0, \\
j_{m}^{(p)}+ER & \text{if } m\neq0\,,
\end{cases} \label{eq:resonanzbedingung}
\end{align}
being valid for $ER\ll 1$ and $E/V_0\ll 1$, where $j_{m}^{(p)}$ denotes the $p$-th zero of $J_m$.

\begin{figure}[htb]
\includegraphics[width=0.45 \textwidth]{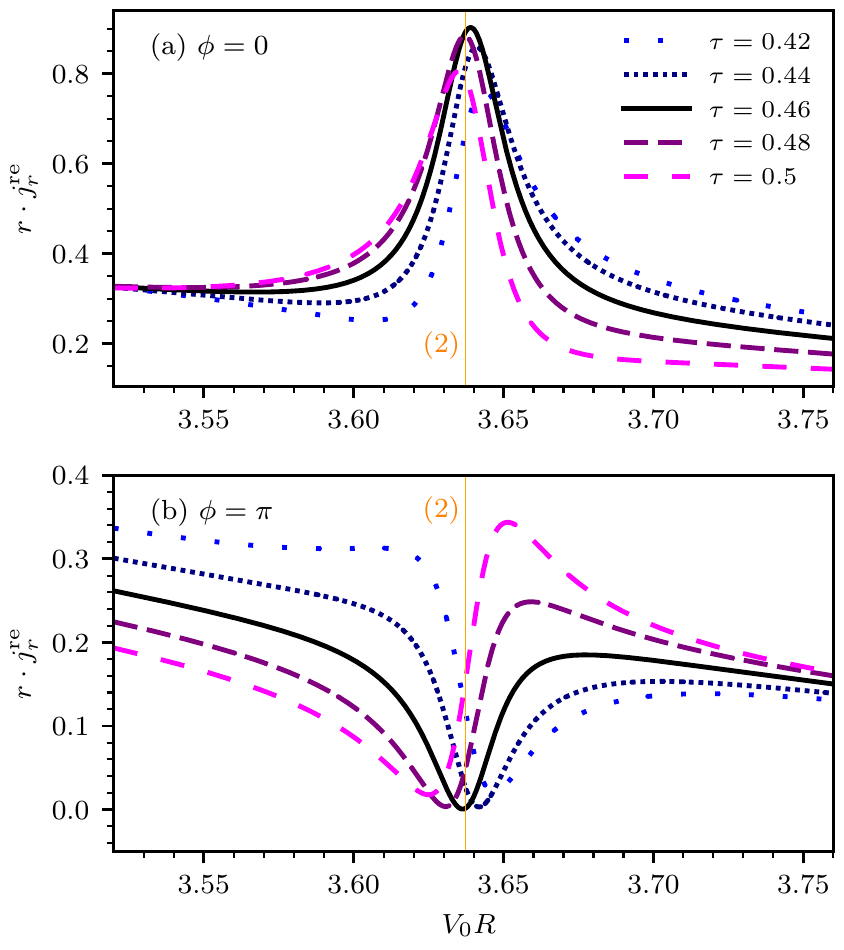}
\caption{Current density of the reflected wave in the far-field as a function of $V_0R$ at different time steps for  $\phi=0$ (a) and  $\phi=\pi$ (b). In both cases model parameters are $E/V_0=0.1$, $V_0=16$, and $\tilde{V}=2$.}
\label{fig:fig6}
\end{figure}

\begin{figure}[htb]
\includegraphics[width=0.45 \textwidth]{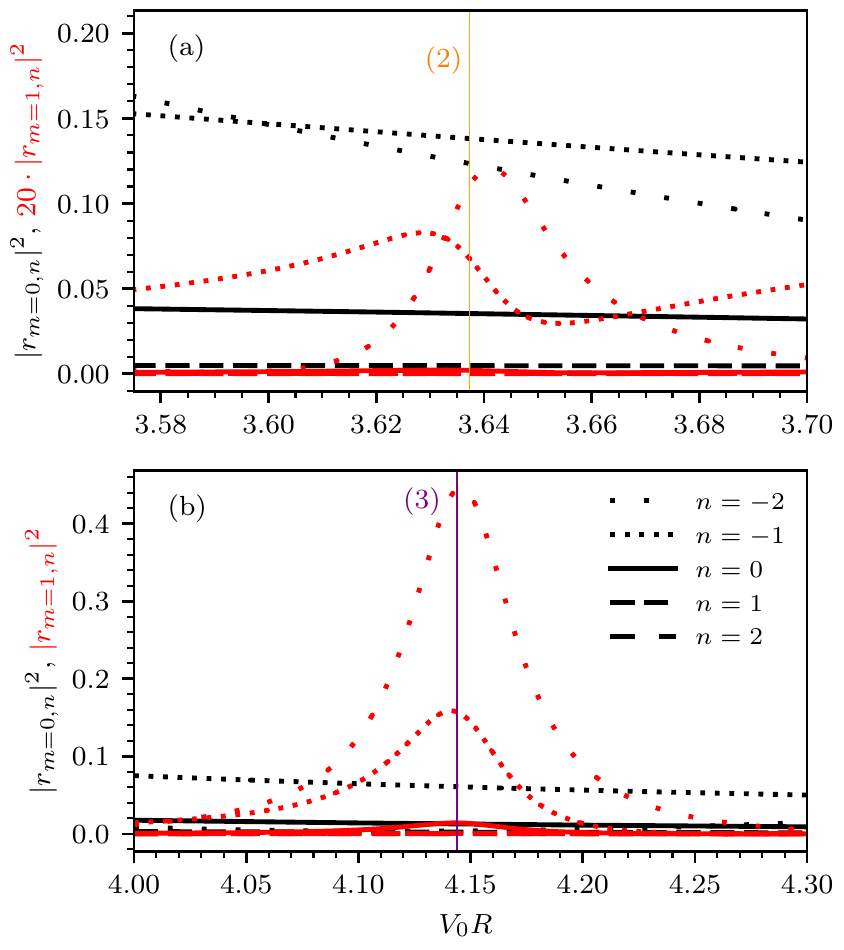}
\caption{Square value of reflection coefficients $|r_{m,n}|^2$ as a function of $V_0R$ near resonances (2) [top panel (a)] and (3) [bottom panel (b)], where the red (black) curves correspond to $m=1$ ($m=0$).  Again, $E/V_0=0.1$ $V_0=16$, and $\tilde{V}=2$.}
\label{fig:fig7}
\end{figure}

An {\it oscillating} quantum dot effectuates significant changes in the scattering behavior through the ability of sideband excitations with $E_n=E+n\omega$. Accordingly the scattering regimes are determined by $E_nR$ and $E_n/V_0$, where the number of sidebands that are relevant increases in relation to $\tilde{V}/\omega$. That is, the static limiting case is obtained for  $\tilde{V}/\omega\to 0$. Furthermore, since the equations derived in the preceding  sections are invariant under the transformation $[E, V_0, \tilde{V}, \omega, R^{-1}]\to \gamma  [E, V_0, \tilde{V}, \omega, R^{-1}]$ with $\gamma \in \mathbb{R}$, we also can fix $\omega=1$, working  hereinafter with rescaled dimensionless parameters.  If we finally want make contact with real world Dice systems, having lattice constants $a$ of about 0.15~nm and nearest-neighbor transfer amplitudes of about 3~eV, the resonant scattering regime ($E/V_0\ll 1$) is realized for quantum dot radii of the order of 100~nm, using oscillation frequencies in the THz regime. 

Figure~\ref{fig:fig3} shows the behavior of the time-averaged scattering efficiency $\overline{Q}$ as a function of $V_0R$ in the resonant scattering regime. 
Panel (a) compares $\overline{Q}$ with the result for the static case. Here we have fixed $E/V_0=0.1$, $V_0=16$ ($V_0$ acts in view of $E_n/V_0 = E/V_0+n/V_0$ as an inverse frequency, i.e., the static behavior is reproduced for very large values of $V_0$ only), and  $\tilde{V}=2$. For these parameters,  in the static case,  some of the resonances given by Eq.~\eqref{eq:resonanzbedingung} are  already smeared out (because a significant number of partial waves contributes to the scattering) while others remain sharp, e.g., at $V_0R\simeq 6$, see the blue dashed line displaying $Q^{\rm static}$. The oscillating dot causes a series of new resonances. For example, resonances (2) and (3) originate from sideband excitations with $n=-1$ and $-2$;  these resonances merge with the $m=1$ resonance of the static case if $V_0\to\infty$.  Of course, the static scattering behavior of the Dice quantum dot will be also rediscovered for $\tilde{V}\to 0$. This general behavior is illustrated by the intensity plots (b) and (c), presenting $\overline{Q}$ in the $V_0R-V_0$ and  $V_0R-\tilde{V}$ planes at fixed $\tilde{V}$ and $V_0$, respectively.  We note that the number and strength of the sideband excitations strongly depends on the amplitude of the oscillations $\tilde{V}$ (for $\tilde{V}=2$ only the   $n=\pm 2, \pm 1$  sidebands give a significant contribution). Varying $V_0$,  the resonance position $V_0R$ is shifted,   which opens up the possibility to tune the interference of different partial waves and sidebands. 
 
\begin{figure}[t]
\centering
\includegraphics[width=0.45 \textwidth]{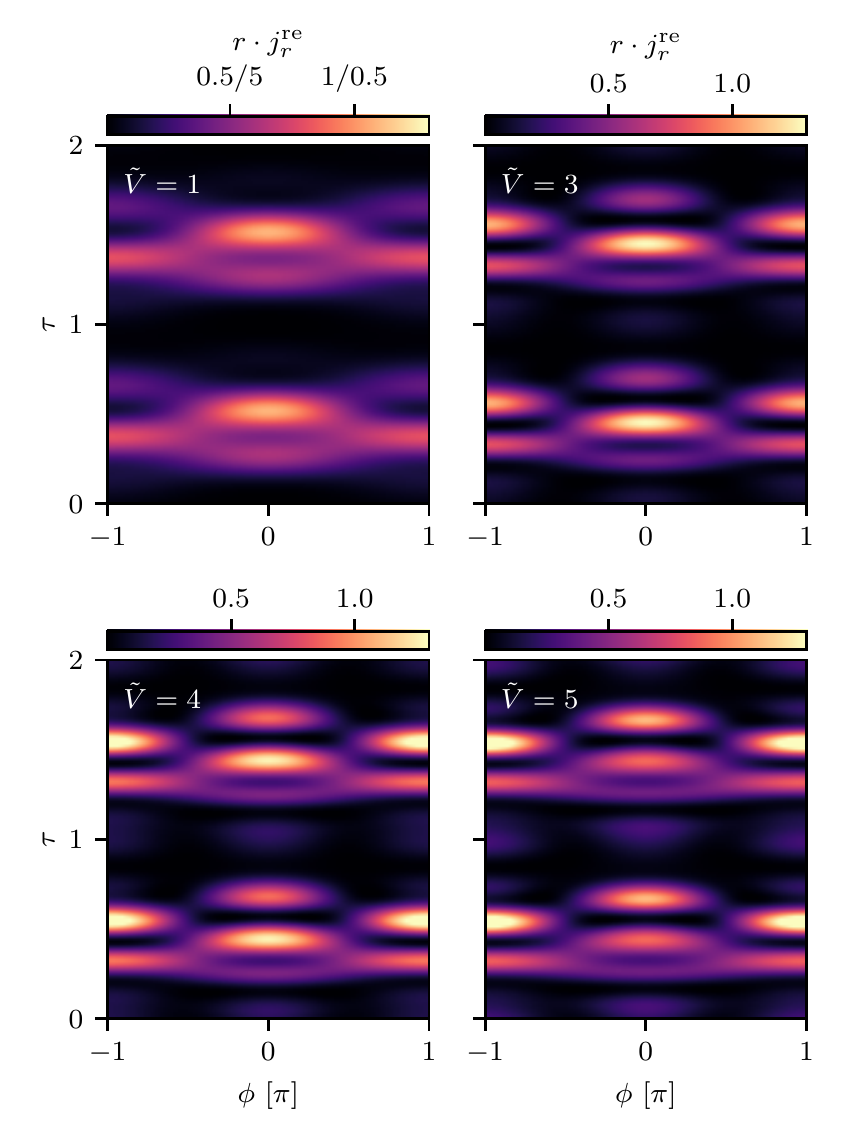}
\caption{Time- and angle-dependent current density of the reflected wave in the far-field, $r\cdot j_r^{\text{re}}(r,\phi,t)$, at $E/V_0=0.1$, $V_0=16$, $V_0R=3.6373$, i.e., at resonance (2), for oscillation amplitudes $\tilde{V}=1$, 3, 4, and 5.}
\label{fig:fig8}
\end{figure}
 Particularly important for potential applications is the time- and/or angle-dependence of the current density in the far-field. Here the radial part, given by Eq.~\eqref{eq:full_ff_curr},  is structured by two Fourier series in the angle and time variables, reflecting the current density and scattering efficiency, respectively,  and a coupling term between them, which---being proportional to $r_{mn}r_{lp}^*$---describes the correlations between the various partial waves in different sidebands. Figure~\ref{fig:fig4} depicts the intensity of $r\cdot j_r^{\text{re}}(r,\phi,t)$ [the factor $r$ just compensates the leading $1/r$-dependence of  $j_r^{\text{re}}(r,\phi,t)$], as function of polar angle $\phi$ and ``time" $\tau=(r-t)/2\pi$, at  specific resonance points $V_0R\simeq 2.65$ (1), 3.64 (2), 4.14 (3), 6.07 (4), 6.72 (5), and 8.10 (6) [cf. Fig.~\ref{fig:fig3} (a)]. All pattern show clearly a $2\pi$-periodicity.  Panel~(1) reveals an almost isotropic scattering characteristics for the Dice quantum dot at low energies, quite contrary to what is observed for oscillating  graphene quantum dots where forward scattering always dominates (cf. Fig.3 in~\cite{SHF15b}). This can be understood as follows. First we can assume that all scattering coefficients with $m\neq 0$ are negligible for $E/V_0=0.1$, $V_0R=2.65$ and $\tilde{V}=2$. Second, considering time-reversal symmetry for  the backscattering process,  the pseudospins of two particles paths that interfere are rotated by $2\pi$ and the phase difference is determined by the Berry phase $\Phi_{\rm B}$ which vanishes for the Dice lattice (but not for a graphene setup where $\Phi_{\rm B}=\pi$). That means both states can interfere in a coherent way, which gives rise to isotropic scattering (in the static case, we find $j_{r=\infty}^{\text{re}}=2v_{\rm F}|r_0|^2/\pi E$). Thereby the incident Dirac pseudospin-1 particle-wave is temporarily captured by the quantum dot and in the sequel, as a result of the dot's potential  oscillation, periodically reemitted at times being multiples of $T=2\pi$ (let us remind that $\omega=1$). Since  higher order partial waves will significantly contribute  to the scattering at larger $V_0R$, a much more complicated  angular and temporal dependence of the radiant emittance develops. This is illustrated by panels (2)--(6). Thus, for example, panel (3) shows the periodic excitation of a  mixed (strong) $m=1$  and  (weak) $m=0$ resonance [where $n=-2$, cf. discussion of Fig.~\ref{fig:fig7} (b) below]. In panel (4)-(6) partial waves with $m$ up to 3 come into play.  The perhaps most interesting scenario develops in panel (2) however: Here forward-scattering and backward-scattering alternate in time, i.e., such a Dice quantum dot can be used for creating a temporal direction-dependent signal  (which is impossible to realize with graphene quantum dots where backscattering is strictly forbidden).  

This ``operational mode''  is displayed in a different way in Fig.~\ref{fig:fig5}.  Here, the radial plots on the left demonstrate how the far-field radiation of the Dice quantum dot is mainly directed forwards or backwards in the course of time, with maximum amplitude at about $\tau=0.46$. The panels on the right give the  current field in the near-field, revealing the characteristics of a (dominant) $m=0$ mode superimposed by a (weak) $m=1$ excitation with two vortices into which the incident wave is fed to. Note that in general the vortex pattern of an $m$-mode is dominated by $2(2m+1)$ vortices located close to the boundary of the quantum dot. This means that the particle is ``confined'' in vortices and not by internal reflections~\cite{BTB09,HBF13a}.  Since the vortices change their orientation in time, the current is driven back and forth through the quantum dot.

The enhancement [suppression] of the far-field reflected current density at $\phi=0$ [$\phi=\pi$] , when passing through resonance (2)  is shown in Fig.~\ref{fig:fig6}~(a) [Fig.~\ref{fig:fig6}~(b)] for the time interval $\tau=0.42\ldots0.5$. Obviously, the constructive and destructive interference between the sharp resonant $m=1$ mode and the broad off-resonant $m=0$ mode can give rise to Fano resonance effects~\cite{Fan61,HBF13a,ZC17}, see, e.g., the result for $\tau=0.42$ [$\tau=0.48, 0.5$] below [above] the point (2) in panel (a) [panel (b)].

\begin{figure}[t]
\centering
\includegraphics[width=0.4 \textwidth]{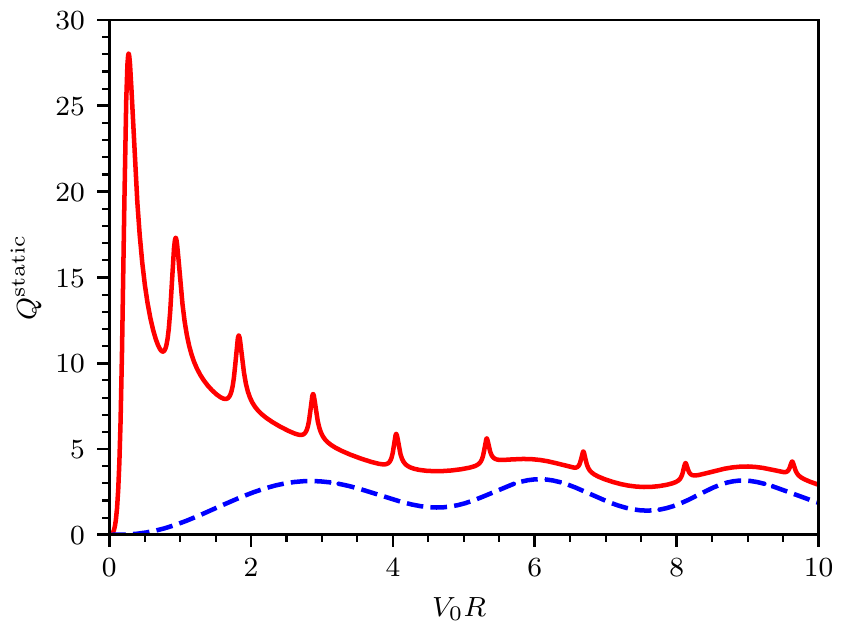}
\caption{Scattering efficiency for the pseudospin-1 Dice model particle (red solid line) compared to the pseudospin-1/2 (graphene model) case (blue dashed line) at $E/V_0=0.49$. }
\label{fig:fig9}
\end{figure}

Figure~\ref{fig:fig7} displays the behavior of the reflection coefficients $|r_{m,n}|^2$ according to Eq.~\eqref{eq:r_mn}. It results from the $m=0, 1$ modes and sidebands $n=-2, -1, 0, 1, 2$ (all other contributions can be neglected). In all cases the $m=0$ mode gives rise to a broad, rather unstructured background scattering only. In the vicinity of resonance (2) [panel (a)], the coupling between the extremely weak (be aware of the scale factor 20 on the ordinate), nearly antisymmetric $n=-1$  and symmetric $n=-2$, $m=1$ modes and the strong $m=0$ mode leads to the rather complex angle- and time- dependent scattering observed in the previous figures.  We note that the contribution to the scattering efficiency is nevertheless significant because in Eq.~\eqref{eq:Q-time} the denominator is proportional to $k_n$, i.e., to $E_\alpha^n$, and might become small if $n<0$. By contrast, in the vicinity of resonance (3) [see panel (b)], the contribution of the $m=0$ mode is almost negligible, and the scattering behavior of the Dice quantum dot is dominated by the $m=1$ mode. As a result, we observe a simultaneous (temporally recurring) emission in both forward and backward directions [cf. Fig~\ref{fig:fig4}(3)]

The influence of the amplitude  $\tilde{V}$ on the reflected current density is illustrated by Fig.~\ref{fig:fig8}, again at resonance~(2). The overall of structure of Fig.~\ref{fig:fig4} (2) (where $\tilde{V}=2$) remains basically unchanged for $\tilde{V}=3$, 4, and 5, only the intensity of the forward scattering shifts from the middle to the later bump in time.  The situation is different for $\tilde{V}=1$. Here, the amplitude seems to be too small to excite the necessary sidebands. As a result the pattern is less structurized and the quantum dot mainly radiates in forward direction.

\begin{figure}[htb]
\includegraphics[width=0.45 \textwidth]{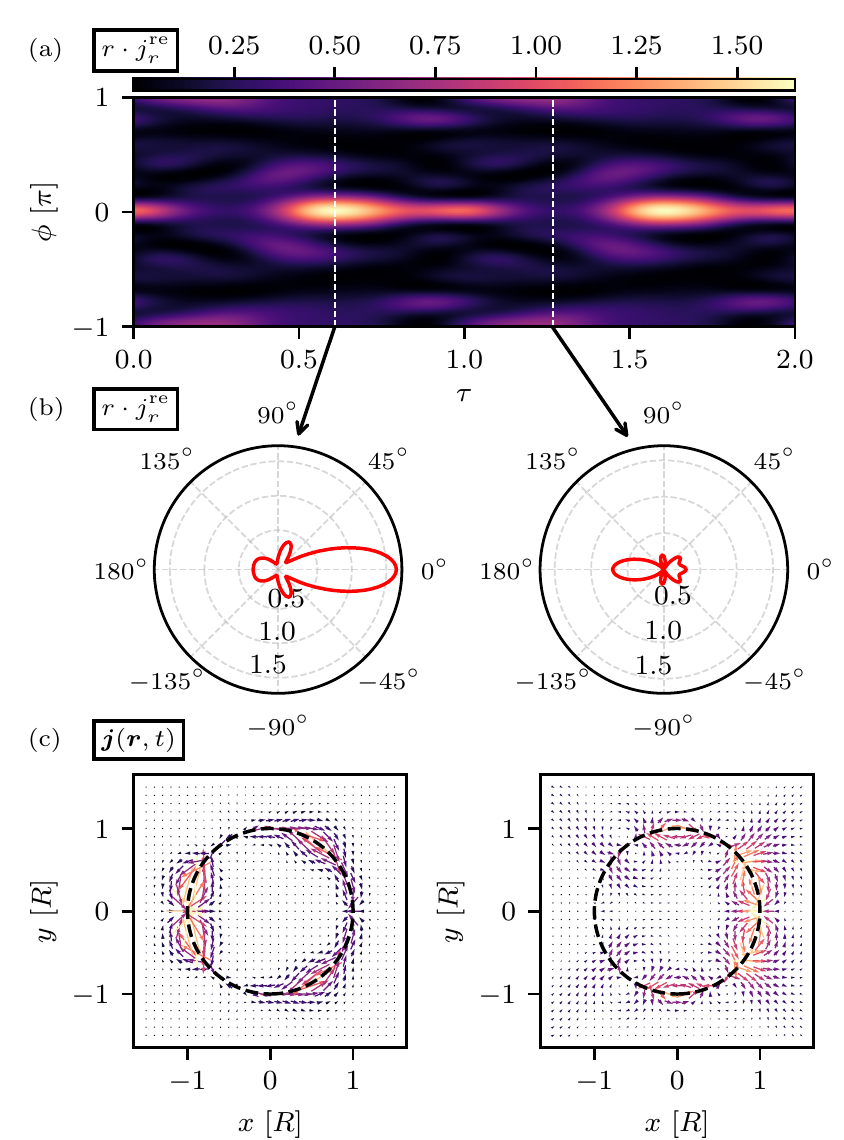}
\caption{
Revival of resonant scattering at $E/V_0 = 0.7$, $V_0R=3.915$, $V_0 = 16$, and $\tilde{V} = 3.06$. (a): Time- and angle-dependent current density of the reflected wave in the far-field, $r\cdot j_r^{\text{re}}(r,\phi,t)$. (b): Related polar plots at the time steps $\tau=0.61$ (left)  and  $\tau=1.27$ (right), indicated in (a) by white dashed lines. (c): Current density in the near-field  at the same time steps. Here, the arrow lengths and colors are given relative to the maximum absolute values of  ${\bf j}({\bf r},t)$, which are 37.73 (left) and 6.11 (right).}
\label{fig:fig10}
\end{figure}

Finally we like to scrutinize the  {\it revival of resonant scattering} for the oscillating Dice quantum dot. The revival of resonant modes has been observed for the  static model only in the vicinity of $E/V_0=0.5$~\cite{XL16}, i.e., in a small window the middle of  the strong reflector regime. Figure~\ref{fig:fig9} compares the behavior of the scattering efficiency $Q^\mathrm{static}$ as a function of $V_0R$ for the Dice model barrier with those for the graphene based one. At $E/V_0=0.49$, we observe for the pseudospin-1 particle a series of sharp resonances (like those found very for small values of $E/V_0$ in the resonant scattering regime), whereas the scattering of the pseudospin-1/2 particle leads to a wavy structure. This effect can be attributed to the formation of specific vortices which are locally attached to the quantum dot boundary. Since these vortices are caused by  wave interference effects, they are very different in nature from whispering gallery modes, and therefore can appear also for small quantum dots~\cite{XL16}. 
\begin{figure}[htb]
\includegraphics[width=0.45 \textwidth]{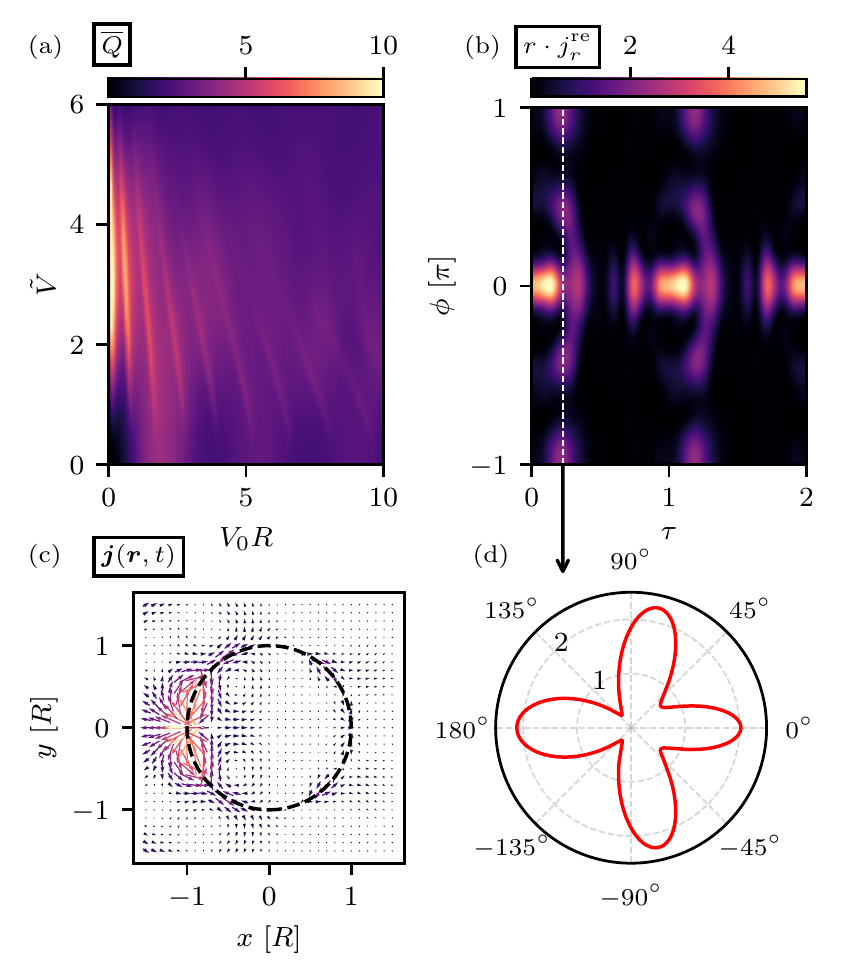}
\caption{Revival of resonant scattering at $E/V_0 = 0.7$ and $V_0=4.76$.  (a): time-averaged scattering efficiency $\overline{Q}$ in the $V_0R$-$\tilde{V}$ plane. (b) 
time- and angle-dependent reflected current density in the far-field, $r\cdot j_r^\text{re}$, where $V_0 R = 3.04$ and $\tilde{V}=4.76$. (c): Corresponding current density in the near-field,  ${\bf j}({\bf r}, t)$, at $\tau=0.23$ [marked in (b) by the white dashed line]. Here,  the arrow lengths and colors are given relative to its maximum absolute value $8.21$. (d) far-field polar plot of $r\cdot j_r^\text{re}$ at the same time step.}
\label{fig:fig11}
\end{figure}

The time-dependent radiation pattern at $E/V_0=0.49$  is depicted in Fig.~\ref{fig:fig10}. Again we find an alternating forward and backward scattering, which is even more strongly directionally focused  than in the normal resonant scattering regime where $E/V_0$ is small [see panels (a) and (b)]. Because of the larger energy of the incident wave, now more partial waves (i.e., higher values of $m$) are involved in the scattering process and the system radiates in other selected directions too, albeit with very small amplitude, see panels (a) and (b).  The reflected current density in the near-field  indicates  that the incident wave is temporally fed into specific trapping profiles, which are strongly shaped by so-called fusiform vortices~\cite{XL16} located at the quantum dot boundary [cf. panels (c)]. 

Interestingly, for periodically driven Dice quantum dots, the revival of resonant scattering will take place in a much wider parameter regime, notably due to the possibility of mixing different scattering regimes.  To demonstrate this, we start out from a quantum dot realizing a strong (Veselago) reflector (with negative refractive index) in the static limit.  Figure~\ref{fig:fig11} exemplarily shows the situation for  $E/V_0 = 0.7$ with $V_0=4.76$. The time-averaged scattering efficiency, reflecting  the strong reflector behavior in the limit  $\tilde{V}\to 0$~\cite{WF14,WF18}, develops the  characteristic resonance pattern as $\tilde{V}$ increases (cf. Fig.~\ref{fig:fig9}), which disappears at large $\tilde{V}$  because now multiple sidebands and partial waves strongly interfere, see panel (a). As regards the time-dependency of the Dice dot emittance, forward scattering is surely dominant but one finds, at certain moments in time, significant radiation in other directions too, see panels~(b) and~(d) for the angle-dependence of the reflected far-field current density.  But the most important finding is the revival of resonance scattering even for large energy potential-barrier ratios, leading to an unexpected strong backscattering. The data presented in panel~(c) for the near-field current density corroborates this statement; see the  pronounced boundary mode (vortex pattern) on the left-hand side.

\section{Conclusions}
 \label{section4}
Physical systems with Dirac cone band structure subject to the action of external temporally varying fields put forward fascinating concepts for novel  electronic and optoelectronic devices. Against this background, we have investigated the time-dependent scattering of a massless pseudospin-1 particle by an oscillating  gate-defined quantum dot placed on the two-dimensional Dice lattice. In the low-energy sector, this setup can be described by an effective Dirac-Weyl theory.  Provided the spatial extensions of the gated region and the wavelength of the Dirac-Weyl  quasiparticles are on the same scale, the quantum resonant scattering regime is realized, where resonances and interference effects play a central role. In experiments this can be achieved with topgates that operate in the terahertz range on nanoribbons. Most importantly, from a theoretical point of view, such oscillating potentials could cause inelastic scattering processes, leading to sideband transitions with $E+n\hbar\omega$ and  $n \in \mathbb{Z}$, which do not exist in the previously studied static barrier setups~\cite{XL16,IN17}. As a consequence, we observed  that the quantum dot traps and emits Dirac-Weyl particle waves periodically in time if certain resonance conditions are fulfilled. Interestingly we could even demonstrate alternating forward and backward scattering which, for sure, is highly relevant for special optoelectronic applications. Equally striking is the observation of a revival of the time-dependent resonant scattering not only when the particle's energy is about half the barrier height of the oscillating Dice quantum dot (as for the static dot) but also for much higher energies of the incident wave. Both effects are related to a mixing of quantum and quasiclassical scattering regimes. Let us point out that all the results obtained for oscillating graphene or Dice quantum barriers so far, are based---to the best of our knowledge---on  effective continuum models. Therefore it might be of particular interest to validate these findings by direct numerical simulations of the corresponding lattice Hamiltonians (just as in the case of static barriers), which, however,  
goes beyond what can be done this work. 

Finally it should be noted that the more general $\alpha-\mathcal{T}_3$-model, which in a sense interpolates between the honeycomb lattice of graphene and the Dice lattice considered here, allows for additional valley skew scattering~\cite{XHHL17,HIXLG19}.  Thus, studying time-dependent scattering in this lattice might pave the way for future valleytronics applications, which opens up interesting prospects for forthcoming theoretical investigations as well.

\begin{acknowledgement}
 \label{section5}
 The authors are grateful to R. L. Heinisch for valuable discussions.
\end{acknowledgement}
\section*{Authors contributions}
All authors outlined the scope and the strategy of the paper. The calculation was performed by AF. HF wrote the mansucript which was edited by all authors. 


\begin{thebibliography}{40}

\bibitem{CGPNG09}
A.H. Castro~Neto, F.~Guinea, N.M.R. Peres, K.S. Novoselov, A.K. Geim, Rev. Mod.
  Phys. \textbf{81}, 109 (2009)

\bibitem{HK10}
M.Z. Hazan, C.L. Kane, Rev. Mod. Phys. \textbf{82}, 3045 (2010)

\bibitem{Xu15}
S.Y. Xu, I.~Belopolski, N.~Alidoust, M.~Neupane, G.~Bian, C.~Zhang, R.~Sankar,
  G.~Chang, Z.~Yuan, C.C. Lee et~al., Science \textbf{349}, 613 (2015)

\bibitem{Kl28}
O.~Klein, Z. Phys. \textbf{53}, 157 (1928)

\bibitem{KNG06}
M.I. Katsnelson, K.S. Novoselov, A.K. Geim, Nature Phys. \textbf{2}, 620 (2006)

\bibitem{SHG09}
N.~Stander, B.~Huard, D.~Goldhaber-Gordon, Phys. Rev. Lett. \textbf{102},
  026807 (2009)

\bibitem{YK09}
A.~Young, P.~Kim, Nature Phys. \textbf{5}, 222 (2009)

\bibitem{BUGH09}
D.~Bercioux, D.F. Urban, H.~Grabert, W.~H\"ausler, Phys. Rev. A \textbf{80},
  063603 (2009)

\bibitem{Su86}
B.~Sutherland, Phys. Rev. B \textbf{34}, 5208 (1986)

\bibitem{UBWH11}
D.F. Urban, D.~Bercioux, M.~Wimmer, W.~H\"ausler, Phys. Rev. B \textbf{84},
  115136 (2011)

\bibitem{LAF18}
D.~Leykam, A.~Andreanov, S.~Flach, Advances in Physics: X \textbf{3}, 1473052
  (2018)

\bibitem{Guea12}
F.~G\"uttinger, J.and~Molitor, C.~Stampfer, S.~Schnez, A.~Jacobsen,
  S.~Dr\"oscher, T.~Ihn, K.~Ensslin, Rep. Prog. Phys. \textbf{75}, 126502
  (2012)

\bibitem{CPP07}
J.~Cserti, A.~P\'alyi, C.~P\'eterfalvi, Phys. Rev. Lett. \textbf{99}, 246801
  (2007)

\bibitem{BTB09}
J.H. Bardarson, M.~Titov, P.W. Brouwer, Phys. Rev. Lett. \textbf{102}, 226803
  (2009)

\bibitem{AU13}
M.M. Asmar, S.E. Ulloa, Phys. Rev. B \textbf{87}, 075420 (2013)

\bibitem{HBF13a}
R.L. Heinisch, F.X. Bronold, H.~Fehske, Phys. Rev. B \textbf{87}, 155409 (2013)

\bibitem{SHF15a}
C.~Schulz, R.L. Heinisch, H.~Fehske, Quantum Matter \textbf{4}, 346 (2015)

\bibitem{XL16}
H.Y. Xu, Y.C. Lai, Phys. Rev. B \textbf{94}, 165405 (2016)

\bibitem{WF14}
J.S. Wu, M.M. Fogler, Phys. Rev. B \textbf{90}, 235402 (2014)

\bibitem{Brea19}
B.~Brun, N.~Moreau, S.~Somanchi, V.H. Nguyen, K.~Watanabe, T.~Taniguchi, J.C.
  Charlier, C.~Stampfer, B.~Hackens, Phys. Rev. B \textbf{100}, 041401 (2019)

\bibitem{HA08}
P.~Hewageegana, V.~Apalkov, Phys. Rev. B \textbf{77}, 245426 (2008)

\bibitem{PAS11}
G.~Pal, W.~Apel, L.~Schweitzer, Phys. Rev. B \textbf{84}, 075446 (2011)

\bibitem{PHF13}
A.~Pieper, R.~Heinisch, H.~Fehske, Europhys. Lett. \textbf{104}, 47010 (2013)

\bibitem{PHWF14}
A.~Pieper, R.L. Heinisch, G.~Wellein, H.~Fehske, Phys. Rev. B \textbf{89},
  165121 (2014)

\bibitem{VAW11}
J.Y. Vaishnav, J.Q. Anderson, J.D. Walls, Phys. Rev. B \textbf{83}, 165437
  (2011)

\bibitem{Caea17}
L.~Camilli, J.H. J{\o}rgensen, J.~Tersoff, A.C. Stoot, R.~Balog, A.~Cassidy,
  J.T. Sadowski, P.~B{\o}ggild, L.~Hornek{\ae}r, Nat. Comm. \textbf{8}, 47
  (2017)

\bibitem{FHP15}
H.~Fehske, G.~Hager, A.~Pieper, Phys. Status Solidi B \textbf{252}, 1868 (2015)

\bibitem{CCOWK16}
J.M. Caridad, S.~Connaughton, C.~Ott, H.B. Weber, V.~Krsti\`{c}, Nat. Comm.
  \textbf{7}, 12894 (2016)

\bibitem{SHF15b}
C.~Schulz, R.L. Heinisch, H.~Fehske, Phys. Rev. B \textbf{91}, 045130 (2015)

\bibitem{WF18}
C.~Wurl, H.~Fehske, Phys. Rev. A \textbf{98}, 063812 (2018)

\bibitem{WF17}
C.~Wurl, H.~Fehske, Sci. Reports \textbf{7}, 9811 (2017)

\bibitem{VOVSDCD13}
M.~Vigh, L.~Oroszl\'any, S.~Vajna, P.~San-Jose, G.~D\'avid, J.~Cserti,
  B.~D\'ora, Phys. Rev. B \textbf{88}, 161413 (2013)

\bibitem{BCGC17}
Y.~Betancur-Ocampo, G.~Cordourier-Maruri, V.~Gupta, R.~de~Coss, Phys. Rev. B
  \textbf{96}, 024304 (2017)

\bibitem{IN17}
E.~Illes, E.J. Nicol, Phys. Rev. B \textbf{95}, 235432 (2017)

\bibitem{AU14}
M.M. Asmar, S.E. Ulloa, Phys. Rev. Lett. \textbf{112}, 136602 (2014)

\bibitem{XL19}
H.Y. Xu, Y.C. Lai, Phys. Rev. B \textbf{99}, 235403 (2019)

\bibitem{Fan61}
U.~Fano, Phys. Rev. \textbf{124}, 1866 (1961)

\bibitem{ZC17}
R.~Zhu, C.~Cai, J. Appl. Phys. \textbf{122}, 124302 (2017)

\bibitem{XHHL17}
H.Y. Xu, L.~Huang, D.~Huang, Y.C. Lai, Phys. Rev. B \textbf{96}, 045412 (2017)

\bibitem{HIXLG19}
D.~Huang, A.~Iurov, H.Y. Xu, Y.C. Lai, G.~Gumbs, Phys. Rev. B \textbf{99},
  245412 (2019)

\end{thebibliography}
\bibliographystyle{epj}

\end{document}